\documentclass[12pt,preprint]{aastex}
\usepackage{lscape}

\def\kms{{\rm km}\,{\rm s}^{-1}}

\begin{document}
\title{$v_\perp$ CMD}

\author{Andrew Gould}

\affil{
Department of Astronomy, The Ohio State University, 
140 West 18th Avenue, Columbus, OH 43210\\
gould@astronomy.ohio-state.edu
}

\begin{abstract}
I present an Hipparcos color-magnitude diagram (CMD) that is color-coded
by transverse velocity $v_\perp$.  This illustrates the connection 
between the photometric and kinematic properties of various stellar
populations in a manner that is particularly suitable to introductory
astronomy courses.
\end{abstract}


\section{Introduction
\label{sec:intro}}

Among the many wonderful uses of the Hipparcos catalog \citep{hip} is that
the Hipparcos color-magnitude diagram (CMD) directly demonstrates 
for students the connections between local stars and the more homogeneous
samples found in clusters.  However, since the local population is
a mixture of populations of various ages and metallicities, the Hipparcos
CMD inevitably appears less distinct than cluster CMDs.

In fact, the Hipparcos catalog contains substantial auxiliary information
to distinguish among these populations in the form of proper motions
(and hence transverse velocities $v_\perp$).
Early main-sequence stars, being young, tend to be moving more slowly
than later main-sequence stars, which are predominantly older.  Halo
subdwarfs, which lie below the main-sequence due to their lower
metallicities, also tend to travel quite fast relative to the Sun.
Thick disk stars are intermediate in both metallicity and kinematics.

\section{CMD
\label{sec:vperp}}

The connection between the photometric and kinematic properties of
the local population can be illustrated simply by color coding the
stars on a CMD according to their $v_\perp$.  For this purpose, I
construct the following subset of the Hipparcos catalog.  First, I
demand that the parallax $\pi\geq 5\sigma_\pi$ to ensure that the
absolute magnitudes $M_V$ and the $v_\perp$ are reasonably accurate.
Then, subject to this restriction, I include all stars satisfying
one of the following three criteria
\hfil\break\noindent 1) $V\leq 7.3$
\hfil\break\noindent 2) $\pi\geq 20\,$mas
\hfil\break\noindent 3) $\mu\geq 200\,\rm mas\,yr^{-1}$

The first selection ensures a good sample of luminous stars.  The second
ensures a representative sample of local stars.  The third ensures that
most halo stars observed by Hipparcos will be included.

\begin{figure}
\plotone{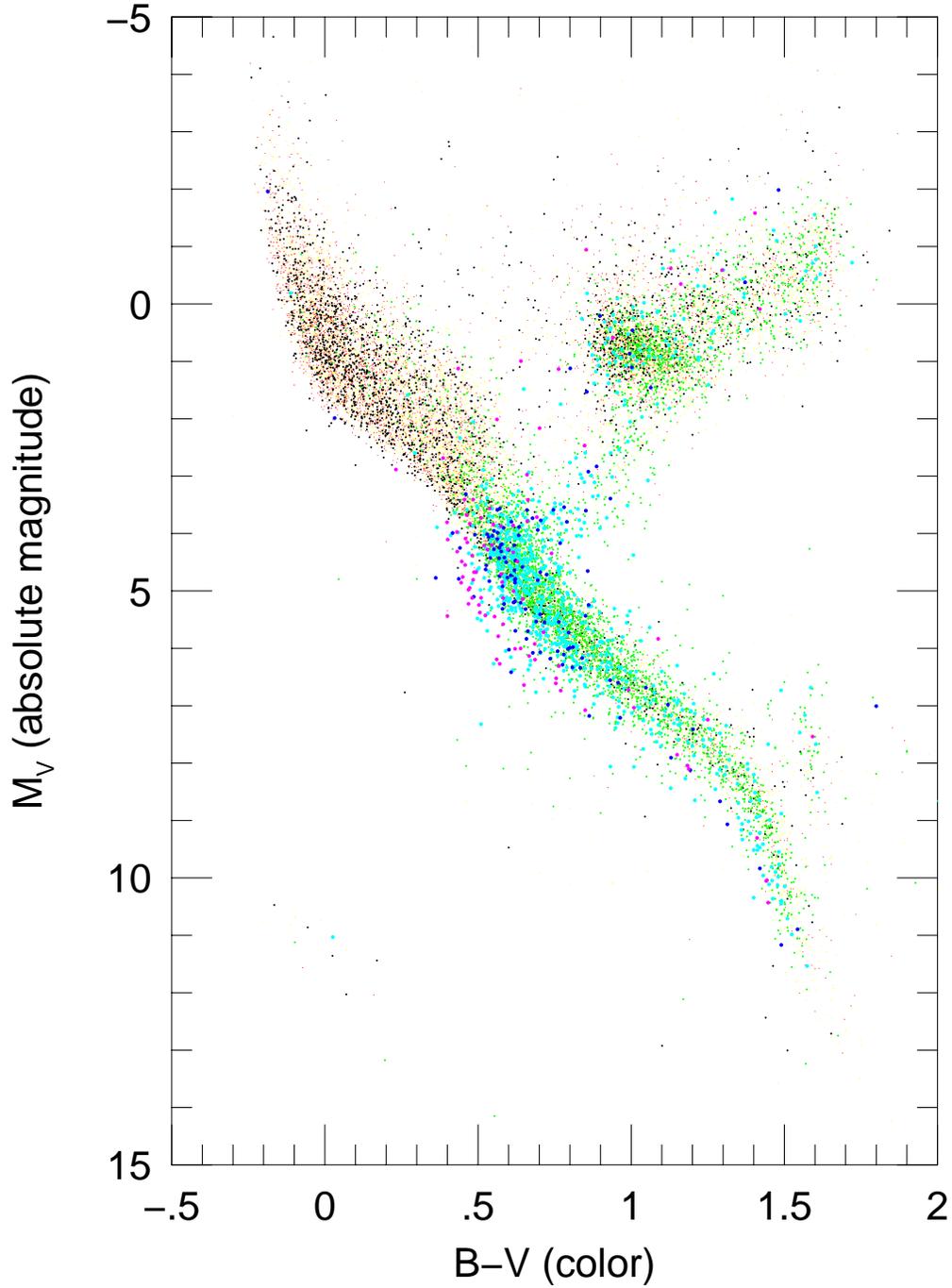}
\caption{Color-magnitude diagram drawn from the Hipparcos catalog
as described in \S~\ref{sec:vperp}.  
 Black: $v_\perp<10\,\kms$;
 Red: $10\,\kms < v_\perp < 20\,\kms$;
 Yellow: $20\,\kms < v_\perp < 50\,\kms$;
 Green: $50\,\kms < v_\perp < 100\,\kms$;
 Cyan: $100\,\kms < v_\perp < 150\,\kms$;
 Blue: $150\,\kms < v_\perp < 200\,\kms$;
 Magenta: $v_\perp > 200\,\kms$.
\label{fig:cmd}}
\end{figure}

Figure \ref{fig:cmd} is the resulting diagram.  The color-coding is:
\hfil\break\noindent Black: $v_\perp<10\,\kms$
\hfil\break\noindent Red: $10\,\kms < v_\perp < 20\,\kms$
\hfil\break\noindent Yellow: $20\,\kms < v_\perp < 50\,\kms$
\hfil\break\noindent Green: $50\,\kms < v_\perp < 100\,\kms$
\hfil\break\noindent Cyan: $100\,\kms < v_\perp < 150\,\kms$
\hfil\break\noindent Blue: $150\,\kms < v_\perp < 200\,\kms$
\hfil\break\noindent Magenta: $v_\perp > 200\,\kms$

\clearpage

\begin{thebibliography}{99}
\frenchspacing

\bibitem[ESA(1997)]{hip} European Space Agency (ESA). 1997, The Hipparcos and
Tycho Catalogues (SP-1200; Noordwijk: ESA)


\end{thebibliography}
\end{document}